\documentclass[a4paper,11pt]{article}
\pdfoutput=1 

\usepackage{jcappub} 
\usepackage{multicol}
\usepackage[T1]{fontenc} 
\usepackage{graphicx}

\title{Understanding Forbush decrease drivers based on shock-only and CME-only models using global signature of February 14, 1978 event}

\author[a,1]{Anil Raghav,\note{Corresponding author.}}
\author[b,1]{Ankush Bhaskar,}
\author[a]{Ajay Lotekar,}
\author[b]{Geeta Vichare,}
\author[b]{and Virendra Yadav}

\affiliation[a]{University Department of Physics, University of Mumbai,\\ Vidyanagari, Santacruz (E),
Mumbai-400098, India}
\affiliation[b]{Indian Institute of Geomagnetism,\\ Kalamboli Highway, New Panvel, Navi Mumbai-
410218, India.}

\emailAdd{raghavanil1984@gmail.com}
\emailAdd{ankushbhaskar@gmail.com}
\emailAdd{ablotekar@gmail.com}
\emailAdd{vicharegeeta@gmail.com}
\emailAdd{virendrahonda@gmail.com}

\abstract{
We have studied Forbush decrease (FD) event occurred on February 14, 1978 using 43 neutron monitor observatories to understand the global signature of FD. We have studied rigidity dependence of shock amplitude and total FD amplitude. We have found almost the same power law index for both shock phase amplitude and total FD amplitude. Local time variation of shock phase amplitude and maximum depression time of FD have been investigated which indicate possible effect of shock/CME orientation. We have analyzed rigidity dependence of time constants of two phase recovery. Time constants of slow component of recovery phase show rigidity dependence and implies possible effect of diffusion. Solar wind speed was observed to be well correlated with slow component of FD recovery phase. This indicates solar wind speed as  possible driver of recovery phase. To investigate the contribution of interplanetary drivers, shock and CME in FD, we have used shock-only and CME-only models. We have applied these models separately to shock phase and main phase amplitudes respectively. This confirms present accepted physical scenario that the first step of FD is due to propagating shock barrier and second step is due to flux rope of CME/magnetic cloud.}

\begin{document}
\maketitle
\flushbottom
\section{Introduction}

\label{sec:intro}

The temporal variation in cosmic radiation at the Earth sometimes shows decrease in its count rate and recovery which typically last for about 8 to 10 days. This was first observed by Forbush, 1937 and Hess $\&$ Demmelmair, 1937 \cite{Hess&Demmelmair1937, Forbush1937}. These short term decreases in the intensity of the galactic cosmic rays at the Earth's surface are called as Forbush decrease (FD).  It was earlier assumed that the variation was caused either directly or indirectly by geomagnetic disturbances. However, it was later shown that the origin of these decreases was in the interplanetary medium \cite {Simpson1954}. These decreases are typically caused by interplanetary counterparts of coronal mass ejections known as `ejecta' and `shock' \cite{cane2000}. The decrease can also be caused by co-rotating interaction regions (CIRs) formed by the fast and slow solar wind streams from the Sun \cite{Lockwood1971}. Majority of FDs show two stpdf during the decrease. Generally, it is believed that the first step is due to the shock and the second is due to the ejecta \cite{wibberenz1998, cane2000, cane1994, cane1996}. It has been reported that  decrease and recovery due to shock are more gradual and symmetric in profile, whereas ejecta have sharp decrease and fast recovery \cite{Iucci1979b, Iucci1979a}. The FD due to shock has typical recovery time of 8 days whereas the decrease due to ejecta has a typical duration of 24 hours \cite{nachkebia2001}. 
The observation of two step or one step FD depends not only on the structure of the interplanetary disturbance but also on the location of the observer. If the Earth crosses only ejecta or shock then it manifests as one step decrease. Also, when the ejecta is less energetic without shock, one step decrease is observed. Hence, broadly, there are three types of FD, which are caused by only shock, only ejecta or combination of shock and ejecta \cite{cane2000}.

It has been a challenge to the scientific community to separate these two parts and find out the principal contributor in FD. To understand the contribution of these two parts in different energy regimes, it is important to carry out detailed investigation of their energy dependence. Its latitudinal and longitudinal (local time) variations and degree of anisotropy are key points to understand the underlying drivers. To examine the effect of these, global study of FD using world wide distribution of neutron monitors is essential. To reduce ambiguity in observations and interpretations, one needs to select a strong FD with clear features of shock and ejecta effect. We have inspected major FDs in last few decades and identified an event occurred on February 14, 1978, which satisfies the above mentioned criteria. We have studied this extraordinary depression using world wide neutron monitor network of 43 observatories. A solar flare occurred at 0129 UT on February 13 at W18, N16 in McMath region no. 15139 producing large amount of solar particles. The average velocity of the shock front estimated was $\sim$ 900 km/s and the maximum velocity of the solar wind associated with this event was $\sim$ 700 km/s. There was a geomagnetic sudden storm commencement (SSC) at 2147 UT on February 14 which was followed by FD \cite{wada1979}. Wada et al. (1979) investigated this event using world wide neutron monitor network and Japanese multi-directional meson telescope. They have examined and reported onset time, sharp decrease and recovery, the time of maximum depression, anomaly in the polar regions and north-south anisotropy. 
Duggal and Pomerantz (1978) reported equator-pole bidirectional anisotropy using seven neutron observatories for the same event. They observed greater magnitude of the FD at the equator as compared to the polar regions and explained the existence of long durational plasma configuration, symmetric with respect to the equator \cite{duggal1978}. Geranios et al.(1983) estimated orientation of the shock by using Helios 1 and 2 and proposed that the western location of the related solar flare might be responsible for the rapid recovery phase \cite{Geranios1983}.

This event is unique due to its distinct two step decrease and two phase recovery. To investigate this unique event in light of the current understanding of FD, we are revisiting this event. This report is arranged as follows, first section discusses the FD characteristics and gives an overview of past works. Details of data and its processing methods are explained in section 2 of the paper. Section 3, 4 and 5 presents various characteristics of the selected FD event, such as rigidity dependence, local time variation, recovery respectively. In Section 6, we briefly discussed CME-only and Shock-only model. finally, we discuss and conclude our observations with respect to current understanding of FD in section 7.
 
\section{Data and methodology}
We have used neutron monitor data from 43 observatories available online at World Data Center in Russia and Ukraine (\url{http://www.wdcb.ru/stp/data/cosmic.ray}) for February 14, 1978 event. We have used Disturbed Storm Time ($D_{st}$) index to determine the shock arrival at ground and  Helios 1 and 2 spacecraft data as proxy for interplanetary conditions (\url{www.cdaweb.gsfc.nasa.gov}). The details of neutron monitor observatories used in the present study are shown in Table  \ref{tab:lab_details}. These observatories are located at different latitudes and longitudes, having different vertical cut-off rigidities (also displayed in the table \ref{tab:lab_details}) which give global representation of FD. Though, FD is a global phenomena, individual observatories will have different baselines due to their respective instrumental characteristics and local parameters. To account for this, we normalized the counts for each observatory by taking the average of 10 previous day counts, which was a quiet period. The normalized percentage variation for each observatory is estimated as follows,   
 
\begin{equation} \label{eq:norm}
 N_{norm} = \frac{N_t - N_{mean}}{N_{mean}} \times 100
\end{equation}

where $N_{mean}$ is average of 10 previous day counts of a particular observatory and $N_{t}$ is neutron count at time `t' for the same observatory. 

\begin{figure}[tbp]
\centering 
\includegraphics[width = 12 cm]{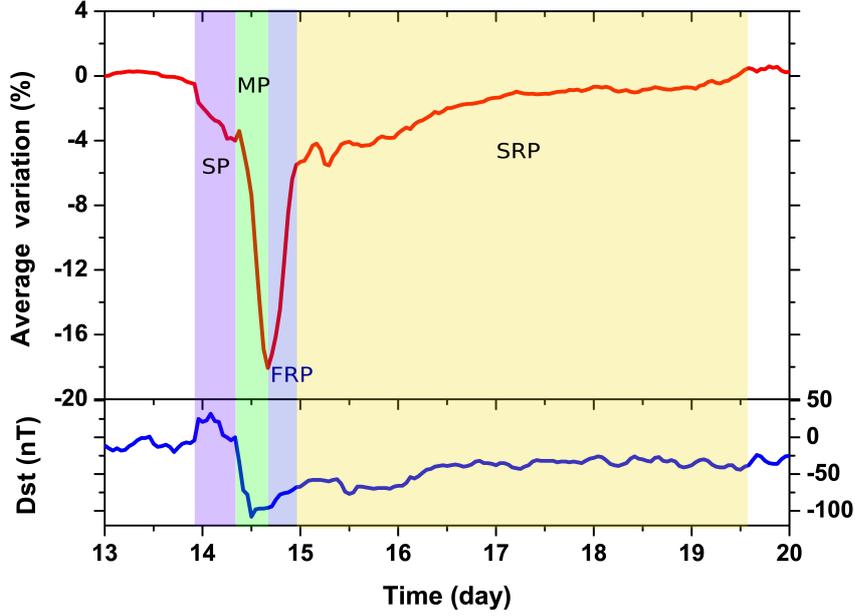}
\caption{Average profile of February 14, 1978 FD. Shaded regions show different phases of FD namely, shock phase (SP), main phase (MP), fast recovery phase (FRP) and slow recovery phase (SRP).  $D_{st}$ index, a proxy of geomagnetic storm, is shown in the lower panel.}
\label{fig:average}
\end{figure}

To understand the average global response of the FD, we have taken the normalized neutron counts of 43 observatories and constructed the average profile, which is shown in figure \ref{fig:average}. As per the current understanding of FD, we can see three distinct phases; shock phase (SP), main phase (MP) and recovery phase (fast (FRP) and slow (SRP)). Shock phase starts with the arrival of shock at magnetopause, which clearly coincides with sharp rise of $D_{st}$ index, termed as sudden storm commencement (SSC). The main phase of the geomagnetic storm reached its maximum intensity prior to the maximum depression of FD. Note that onset of main phase of the storm seen in $D_{st}$ index coincides with onset of main phase of FD and indicates incidence of flux rope at magnetopause. This implies that main phase of FD started when the Earth entered the flux rope. Shock phase of FD shows gradual decrease, whereas main phase of FD shows sharp decrease.  Shock phase lasted for about 12 hours and  main phase duration is about 6 hours. The recovery phase of FD is composed of rapid and gradual components which lasted for about 6 hours and few days respectively. The average peak magnitude of FD is 18 $\%$ in which the shock phase average amplitude is about 4 $\%$. It is important to note that the main phase and rapid component of recovery phase show symmetry in profile, where rapid phase recovers nearly to the minimum depression of shock phase.

We estimated FD amplitude (Total) for individual observatories, which is the maximum depression from the normalized baseline. As stated in section \ref{sec:intro}, the decrease in cosmic ray flux is two step, in which shock phase amplitude of FD is the maximum depression from the normalized baseline, just prior to the sharp decrease. FD maximum depression time is determined as the time when the main phase (MP) of FD ends. Generally, the recovery phase is expressed as a single exponential function. However, the studied event shows two distinct phases in recovery; a rapid and a gradual one.  Therefore, we have fitted double exponential function to complete recovery phase. Since, cosmic ray intensity (neutron counts) is measured at different longitudes, latitudes and with different vertical cut-off rigidities, we have investigated the effect of each parameter by minimizing the effects of other parameters.

\begin{table}[H]
\caption{Details of neutron monitor observatories used in present study} 
\centering

\footnotesize

\begin{tabular}{c c c c c c}
\hline
\hline
Lab Number  &Lab Name  &Latitude  &Longitude  &Rigidity\\[0.5ex]
\hline

1	&AATA	&43.25	&76.92		&6.66\\
2	&AATB	&43.14	&76.6		&6.69\\
3	&APTY	&67.55	&33.33		&0.65\\
4	&CALG	&51.08	&-114.13	&1.08\\
5	&CAPS	&68.92	&-179.47	&0.45\\
6	&CLMX	&39.37	&-106.18	&3.03\\
7	&DPRV	&46.1	&-77.5		&1.02\\
8	&DRBS	&50.1	&4.6		&3.24\\
9	&DRHM	&43.1	&-71		&1.41\\
10	&FUSH	&37.68	&140.45		&10.55\\
11	&GSBY	&53.27	&-60.4		&0.52\\
12	&HRMS	&-34.42	&19.22		&4.9\\
13	&INVK	&68.35	&-133.72	&0.18\\
14	&IRKT	&52.47	&104.03		&3.66\\
15	&01/Jun	&46.55	&7.98		&4.48\\
16	&KERG	&-49.35	&70.25		&1.19\\
17	&KIEL	&54.3	&10.1		&2.29\\
18	&KIEV	&50.72	&30.3		&3.62\\
19	&LEED	&53.8	&-1.5		&2.2\\
20	&LMKS	&49.11	&20.13		&4\\
21	&MGDN	&60.1	&151		&2.1\\
22	&MOSC	&55.47	&37.32		&2.46\\
23	&MTNR	&36.12	&137.55		&11.39\\
24	&MTWL	&42.92	&147.23		&1.89\\
25	&MTWS	&44.28	&-71.3		&1.24\\
26	&NLCH	&43.3	&43.25		&7.7\\
27	&NRLK	&69.26	&88.05		&0.63\\
28	&NVBK	&54.8	&83		&2.91\\
29	&OULU	&65.02	&25.5		&0.81\\
30	&PTFM	&-26.68	&27.92		&7.3\\
31	&ROME	&41.9	&12.5		&6.32\\
32	&SNAE	&-70.3	&-2.35		&1.06\\
33	&SOPO	&-90	&0		&0.11\\
34	&SVER	&56.8	&60.63		&2.3\\
35	&SWTH	&39.9	&-75.35		&1.92\\
36	&TASH	&41.33	&69.62		&8.34\\
37	&TBLS	&41.72	&44.8		&6.91\\
38	&TKYO	&35.75	&139.72		&11.61\\
39	&TSMB	&-19.2	&17.6		&9.29\\
40	&TXBY	&71.6	&128.9		&0.53\\
41	&UTRT	&52.1	&5.12		&2.76\\
42	&YKTK	&62.02	&129.72		&1.7\\
43	&ZUGS	&47.42	&10.98		&4.24\\

\hline 
\end{tabular}
\label{tab:lab_details}
\end{table}

\section{Rigidity dependence of FD}

\begin{figure}[tbp]
\begin{center}
\includegraphics[width = 10 cm, height = 16 cm]{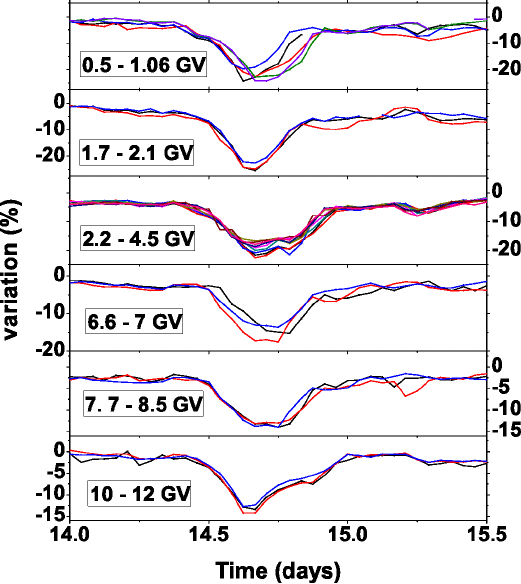}
\caption{Energy dependence of temporal $\%$ variation of neutron flux during February 14, 1978 FD event for for observatories from northern hemisphere, located between $5^{\circ}$E to $152^{\circ}$ E longitudinal band. }
\label{fig:stack_rigi}
\end{center}
\end{figure}

The widely spread neutron monitor observatories having different vertical cut-off rigidities (energies) are analogues to a spectrometer, used to study the spectral response of cosmic rays during interplanetary disturbances. Based on this analogy we have investigated energy dependence of FD evolution. The temporal variations of neutron counts for different ranges of vertical cut-off rigidity are studied and shown in figure \ref{fig:stack_rigi}. To minimize possible longitudinal and hemispherical anisotropy, we have selected observatories from northern hemisphere, located between $5^{\circ}$E to $152^{\circ}$E longitudinal band. The wide longitudinal band is due to the constraint on availability of observatories. Note that the shape of FD profile is different for different ranges of vertical cut-off rigidity. Figure \ref{fig:stack_rigi} clearly shows onset of FD main phase is simultaneous in different energies, whereas the amplitude of FD decreases with increasing rigidity. These selected observatories show approximately the same maximum depression time from the baseline.  Most of the profiles in different energy ranges show extended minimum. It has been clearly observed that below 2.1 GV, rigidity profiles show sharp decrease and sharp recovery. However, above this rigidity, we observed extended minimum which increases with increasing rigidity except 10-12 GV rigidity.

\begin{figure}[tbp]
\begin{center}
\includegraphics[width = 12 cm]{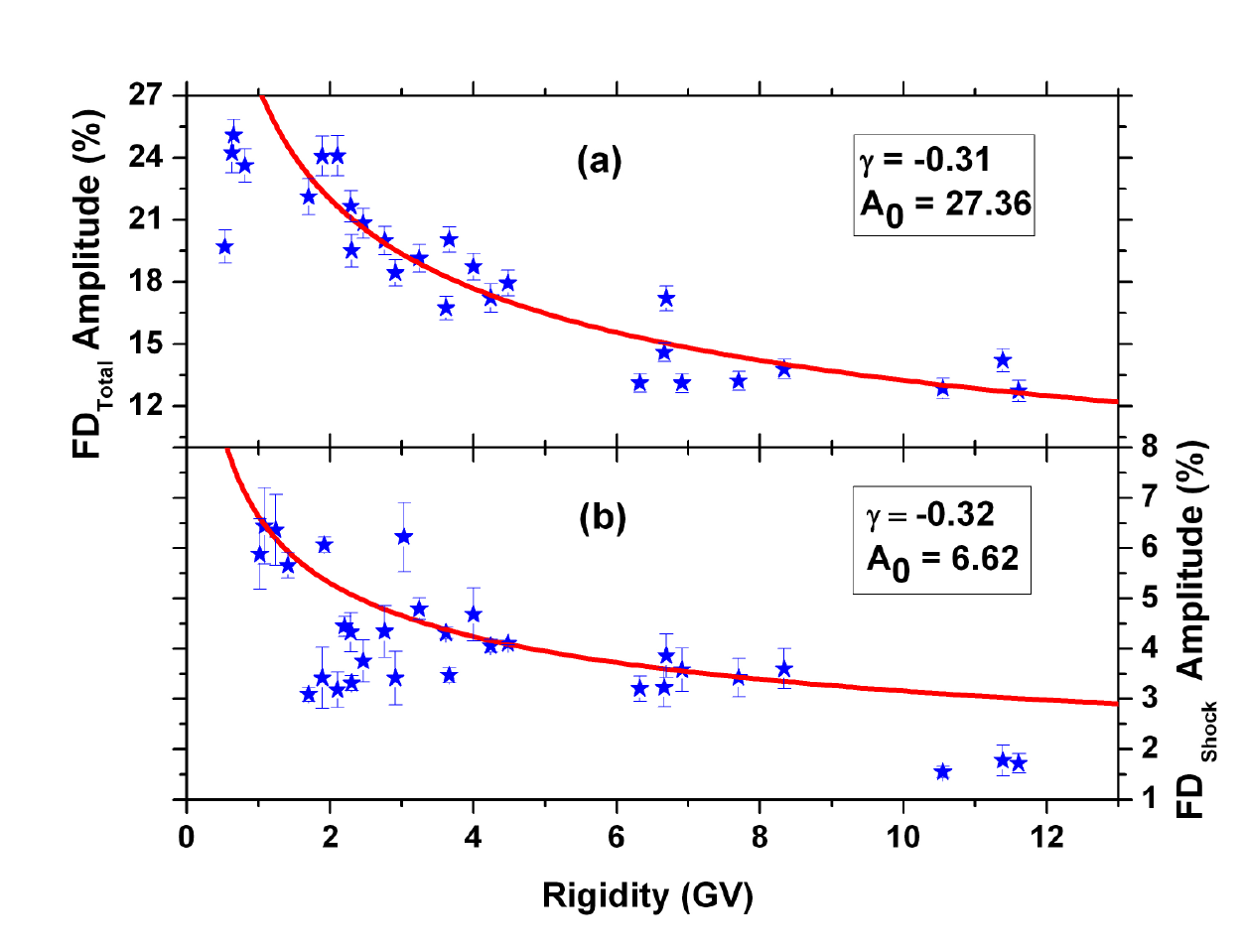}
\caption{Rigidity dependence of total FD amplitude (a) and FD shock phase amplitude (b) for observatories located in northern hemisphere and situated between $35^\circ$ N to $62^\circ$ N and $-106^\circ$ E to $152^\circ$ E having vertical cutoff rigidity  $> 1 GV$.}
\label{fig:rigi_fit}
\end{center}
\end{figure}

Rigidity dependence of FD is quantified by expressing the FD amplitude as function of the vertical cut-off rigidity for the same observatories as shown in figure \ref{fig:rigi_fit}. The observed FD amplitude varies from $\sim$ 25 $\%$ to $\sim$ 13 $\%$ corresponding to rigidities from  $\sim$ 0.5 GV to  $\sim$ 12 GV. It has been reported that the FD amplitude - rigidity relationship exhibits power law \cite{Lockwood1991}.  This power law is presented by the following function,  

\begin{equation}
A = A_0 R^{-\gamma}
\label{eq:power_law}
\end{equation}
 where, A is observed FD amplitude for given rigidity, R is vertical cut-off rigidity, $\gamma$ is power index and $A_0$ is a constant.
 
To estimate the power law index for the rigidity dependence of FD, we have fitted the data with the above mentioned power function by using least square fitting. We have found $\gamma = 0.31$, which is within the range  $0.2 < \gamma < 0.8$ reported by Kane (1963) \cite{kane1963}. The FD amplitude at cut-off rigidity $\sim$ 1 GV  is about twice of that at $\sim$ 12 GV, which is consistent with earlier observation \cite{Lockwood1971}. The power law behaviour of the system indicates the self-similar nature of the FD in the different energy regimes for 0.5 - 12 GV range. Since, the physical mechanisms behind the two phases of FD are different, we have studied the shock phase separately. We have separated shock phase amplitude from the total FD amplitude. The rigidity  dependence of the shock phase amplitude is shown in Figure \ref{fig:rigi_fit}.  The magnitude  of the shock phase also follows a power law behaviour. The estimated power law index is $\gamma = 0.32 $.

\section{Local time variation } \label{subsec:lT}

\begin{figure}[tbp]
\centering
\includegraphics[width = 14 cm]{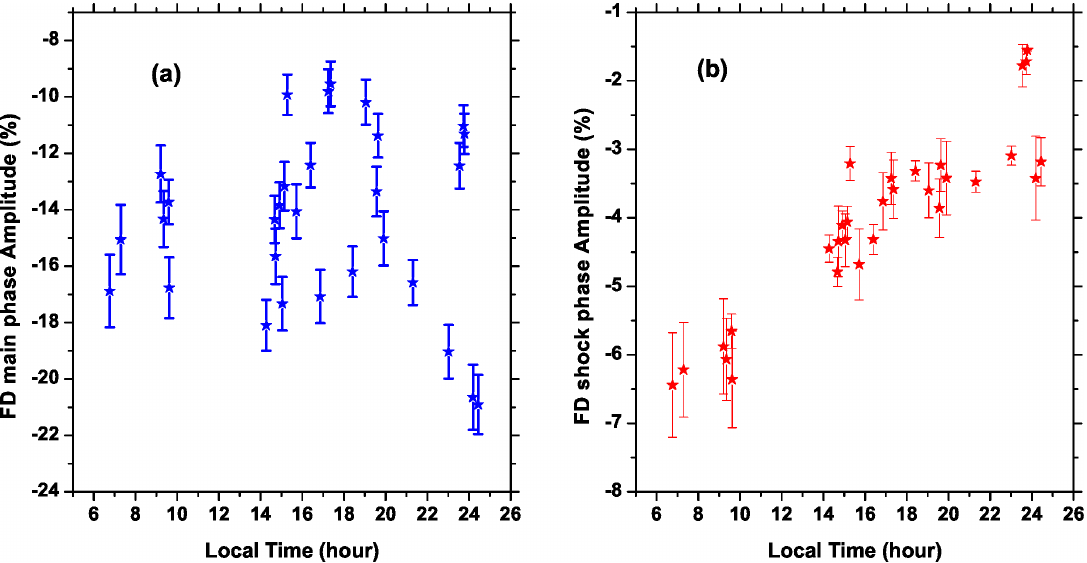}
\caption{Local time variation of main phase (a) and shock phase (b) amplitudes for observatories located in northern hemisphere and situated between $35^\circ$ N to $62^\circ$ N and $-106^\circ$ E to $152^\circ$ E having vertical cutoff rigidity  $> 1 $ GV.}
\label{fig:LT_shock}
\end{figure}

The interplanetary disturbances engulfing the Earth are generally of the order of 1 AU. However, the inhomogeneities and impact direction of interplanetary disturbance might give rise to anisotropies along different longitudes. To unravel this effect, local time (LT) variation of FD should be studied. Therefore, we have investigated LT modulation of FD amplitude. However, no clear relationship is observed. As per current understanding, FD is combined effect of shock and ejecta. Therefore, we have separated shock and ejecta components to study their LT variations. To minimize the amplitude variation due to hemispherical anisotropy, rigidity and latitude, we have shortlisted observatories located in northern hemisphere and situated between $39^\circ$ N to $62^\circ$ N having vertical cutoff rigidity  $< 3.1 $ GV. Once again no clear LT variation of ejecta component (main phase amplitude) is observed. However, The local time modulation of shock phase amplitude is clearly seen as depicted by figure \ref{fig:LT_shock}. The maximum shock phase amplitude is approximately $6.5 \%$ and minimum is approximately $3\%$. Note that, shock phase amplitude is maximum at dawn and continuously decreases towards dusk and afterwards remains nearly constant. 

\begin{figure}[tbp]
\centering
\includegraphics[width = 10 cm]{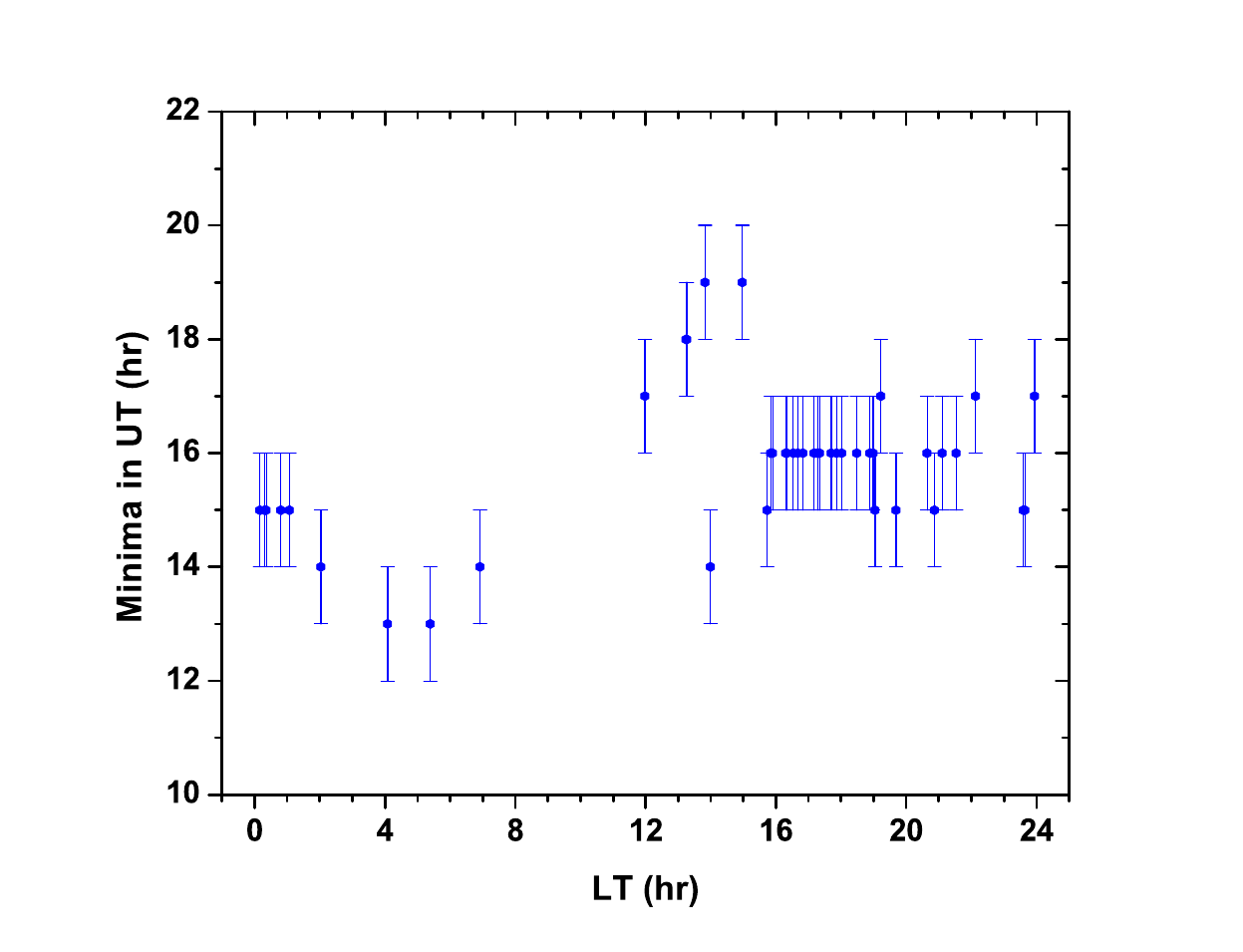}
\caption{Variation of maximum depression time in Universal time (UT) with respect to the local time (LT)}
\label{fig:mini_ut}
\end{figure}

\begin{figure}[tbp]
\centering
\includegraphics[width = 10 cm]{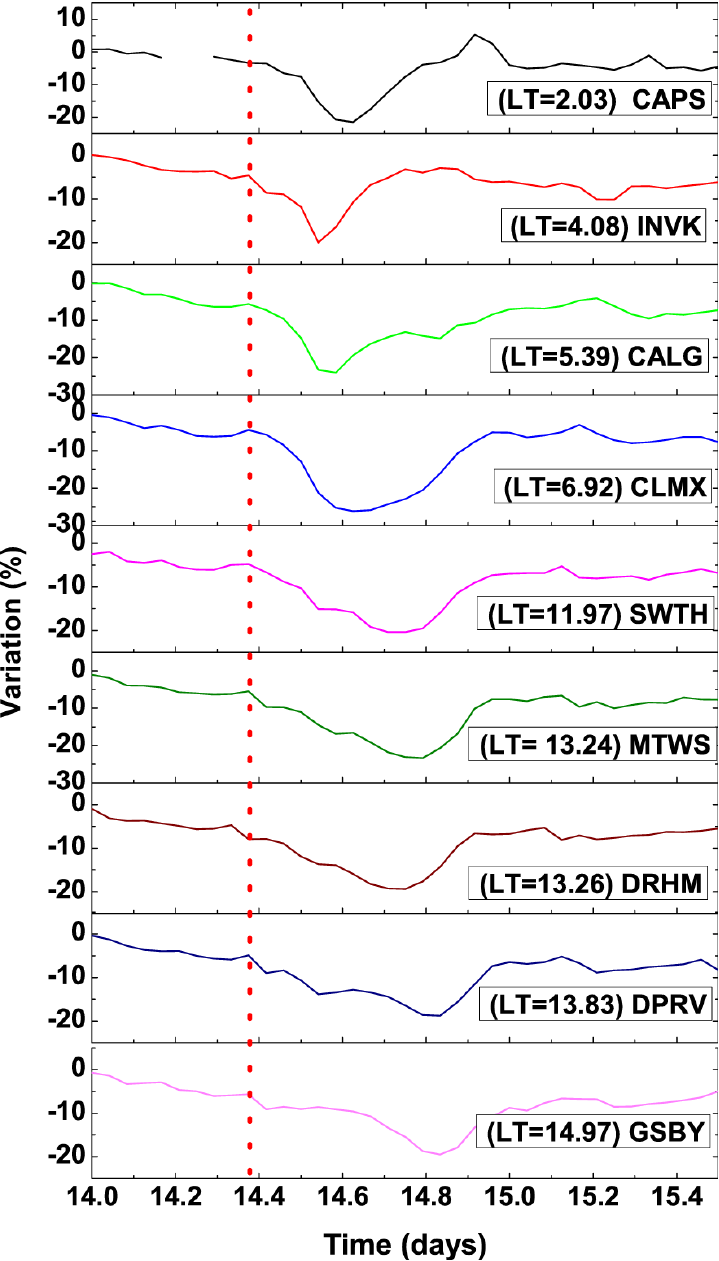}
\caption{Stack plot of normalized neutron flux variation for observatories located in $\sim$ 0200 to 1500 LT region.}
\label{fig:stack_longi}
\end{figure}

We have also attempted to understand the local time effect on time of maximum depression of the FD. This effect is clearly seen in figure \ref{fig:mini_ut}. Here, the LT value for a given observatory is the time of minimum of main phase observed at that observatory.  It is interesting to see that $\sim$ 1600 LT onwards minimum occur almost at same time in UT which is continued till $\sim$ 0200 LT. However, it is observed that minimum occur earlier near dawn as compared to that of afternoon. It is important to note that from $\sim$ 0400 LT onwards time of maximum depression is continuously increasing till $\sim$ 1500 LT and then at $\sim$ 1600 LT returns back to almost same UT time. To investigate details of this rising trend of time of maximum depression, we have shown their respective FD profiles (for $\sim$ 0200 to 1500 LT ) in figure \ref{fig:stack_longi}. It is clearly seen that the profiles of FDs vary with LT. The duration of main phase increases with increase in LT. Also, recovery phase of the FDs shows difference in the duration. The onset time of shock phase is approximately simultaneous at almost all LT's. However, at few observatories shock phase amplitude is too small to detect onset.

\section{Recovery}

\begin{figure}[tbp]
\begin{center}
\includegraphics[width = 10 cm]{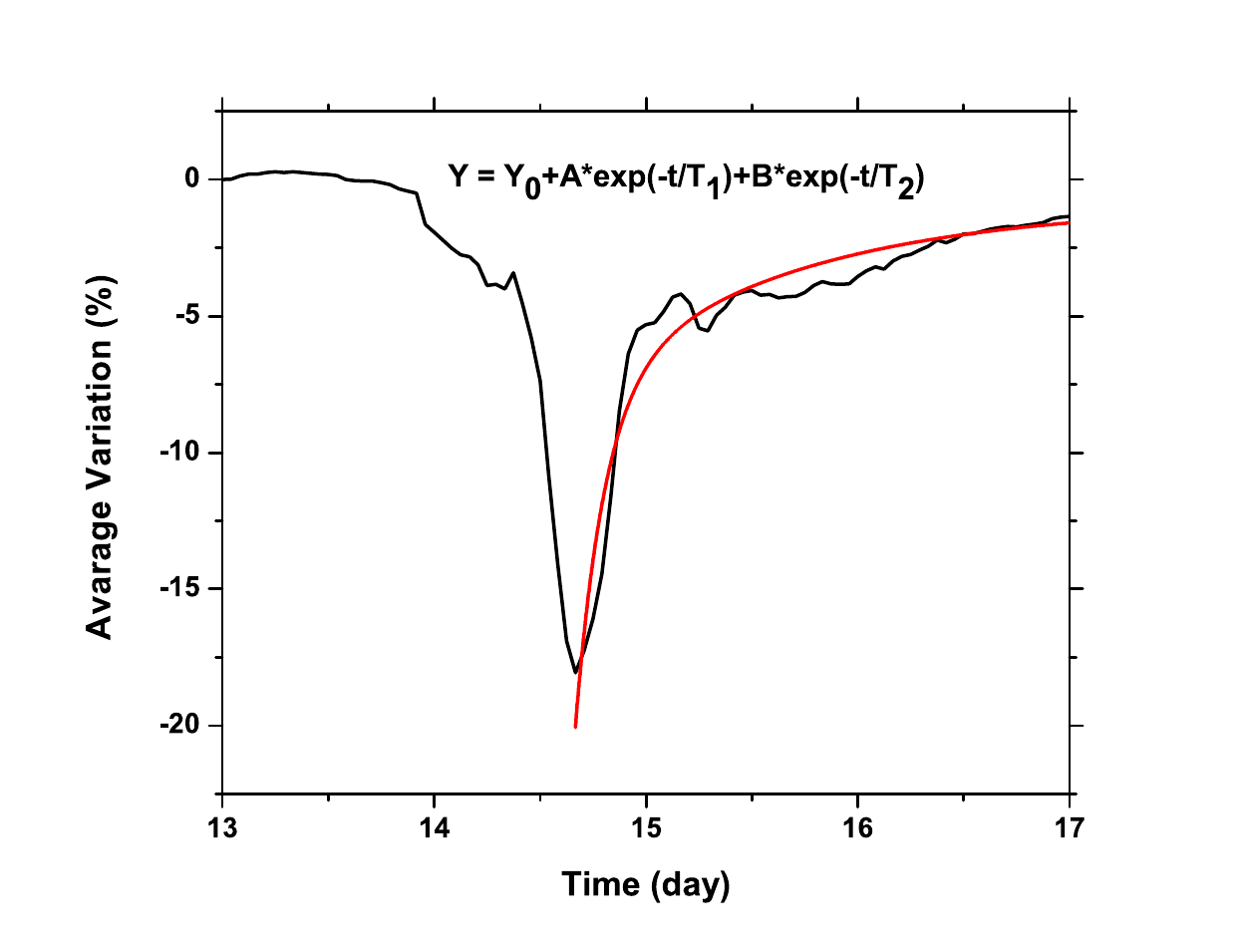}
\caption{Typical double exponential fit to recovery phase of average neutron flux variations for the event.}
\label{fig:dubel_fit}
\end{center}
\end{figure}

\begin{figure}[tbp]
\begin{center}
\includegraphics[width = 10 cm]{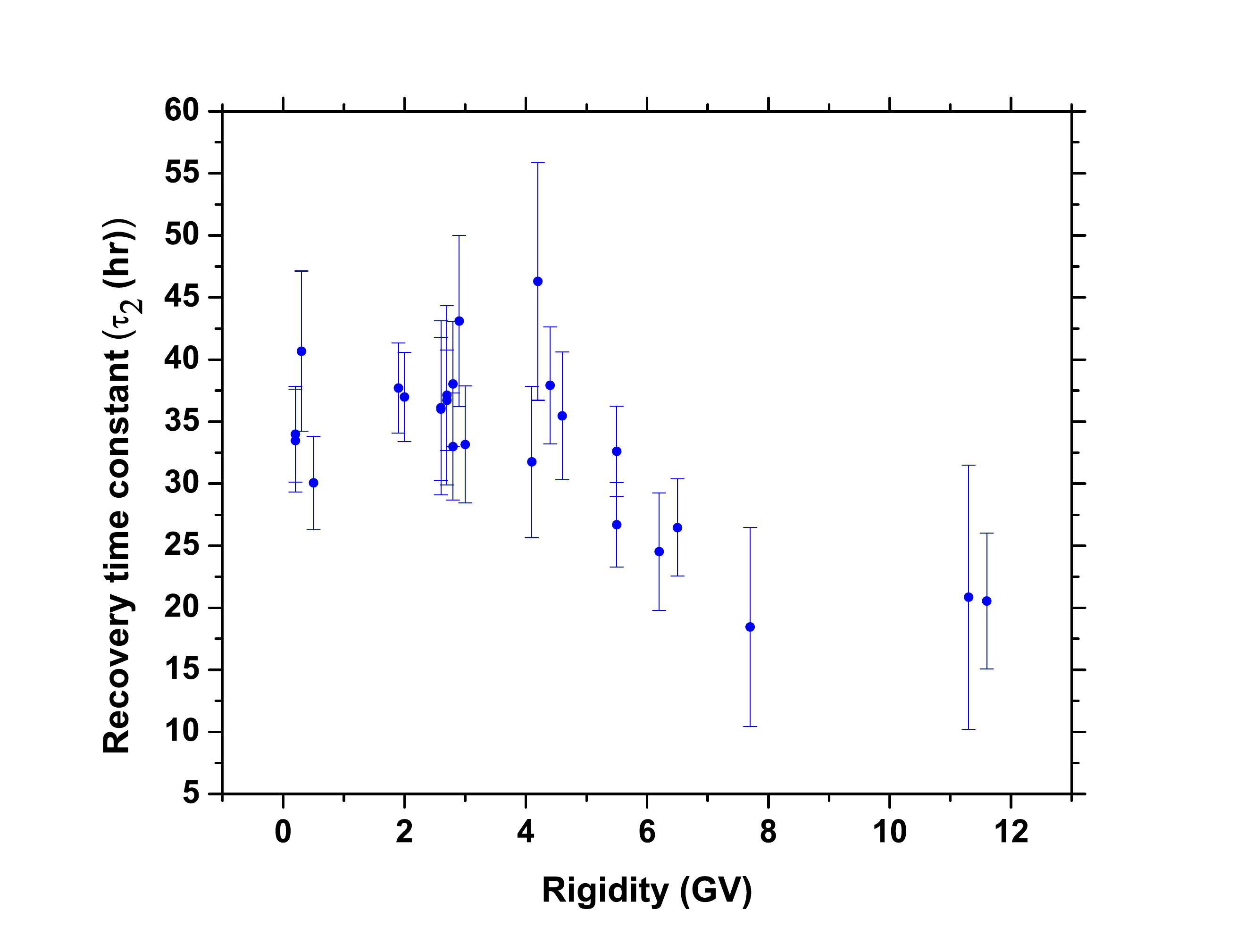}
\caption{Rigidity dependence of time constant ($\tau_2$) for slow phase of recovery.}
\label{fig:rigi_t2}
\end{center}
\end{figure}

The recovery phase of FD is very interesting aspect due to its variability in duration and unique two phase profile. It has been reported in past studies that recovery generally follows an exponential trend. The present event shows two distinct  phases of recovery. Therefore, instead of generally adopted single exponential function fitting, we have fitted double exponential function for each FD profile by using least square fit. This gives two different time constants. A typical fitted recovery is depicted in Figure \ref{fig:dubel_fit}. The first part of recovery shows small time constant corresponding to fast phase of recovery whereas, second phase of recovery shows high value of time constant corresponding to slow phase of recovery. These time constants are referred as $\tau_1$ and $\tau_2$ respectively. We observed that $\tau_1$ does not show any clear dependence either on the cut-off rigidity or local time. However, $\tau_2$ shows clear dependence on the cut-off rigidity as shown in Figure \ref{fig:rigi_t2} whereas, no unambiguous dependence was observed on local time. Note that $\tau_2$ is almost same for 0.5-4 GV and then, it decreases for higher cut-off rigidities (for $ > 4$ GV and $< 12$ GV).

\begin{figure}[tbp]
\includegraphics[width = 12 cm]{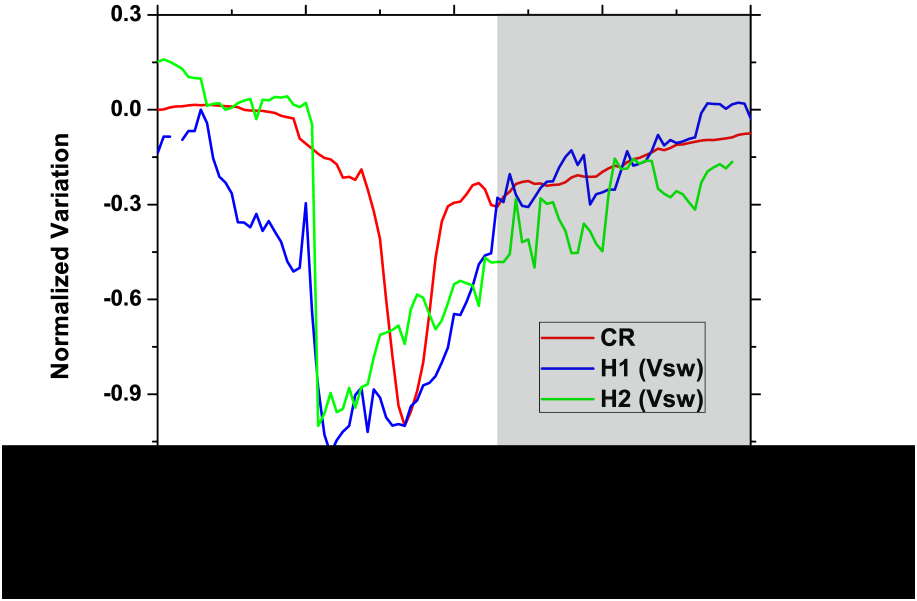}
\caption{The average neutron flux and associated solar wind speed variations observed at Helios 1 and 2. Solar wind speed is inverted and normalized with respect to pre-onset value for better comparison.}
\label{fig:vsw}
\end{figure}

To investigate the influence of solar wind speed on the FD, we have used Helios 1 and 2 spacecraft data.  The solar wind speed ($V_{sw}$) was inverted and normalized with respect pre-onset value to compare it with the FD profile. Figure \ref{fig:vsw} demonstrates both Helios 1 and 2 normalized solar wind datasets and average normalized neutron monitors data. It is observed that the onset of shock phase of FD nearly coincides with sharp rise of solar wind speed observed by both the spacecraft. It is important to note that the slow component of recovery phase of FD (shown by shaded region) correlates well with the corresponding  solar wind speed variations observed by Helios 1 and 2.

\section{Forbush decrease models}

To address the present question in this field, weather FD is due to shock only or CME only or combination of both, we have applied the CME-only and shock-only model \cite{wibberenz1998, wibberenz1997, subramanian2009, Babu_et_al_2013} to the present FD event. The basic formalism of both the models is briefly described here.  The CME-only model assumes that the FD originates only due to the CME. Shock-only model assumes that the FD is exclusively due to the interplanetary shock, approximated as a diffusive barrier. The shock and the CME both are expected to contribute to the FD effectively. Therefore, we have selected a FD event which shows shock and CME component distinctly. Treating shock and ejecta separately is very important aspect of this study since both components contribute in total magnitude of the FD. We believe that Shock phase magnitude of FD can be considered as the maximum contribution due to interplanetary shock in total FD whereas, the difference between total FD magnitude and shock phase magnitude is attributed to the CME only. The main differences between model fitting of earlier work by \cite{subramanian2009, Babu_et_al_2013} and the present work are, (1) we have used global neutron monitor data which give FD variation in different rigidities, (2) shock phase and ejecta phase of FD treated separately.

Both the models are dependent on the diffusion of cosmic rays into the shock or CME which is characterized by diffusion coefficient. The diffusion of the ambient high energy cosmic rays into turbulent CME/magnetic cloud across the magnetic fields that envelope it, is controlled by cross-field diffusion coefficient $D_\perp$. However, Cosmic ray diffusion along the field line is governed by parallel diffusion coefficient $D_\parallel$. Candia and Roulet (2004) have determined perpendicular and parallel diffusion coefficients for charged particles in highly turbulent magnetic fields \cite{candia2004}. They expressed these diffusion coefficients in terms of magnetic rigidity and turbulence level for different types of turbulence spectra. Here, it is assumed turbulence follows Kolmogorov turbulence spectrum \cite{subramanian2009}. 
The parallel diffusion coefficient is given as,
\begin{equation}
D_\parallel = c \rho L_{max}  \frac{N_\parallel}{\sigma^2} \sqrt{ \left({\frac{\rho}{\rho_\parallel}}\right)^{2(1-\gamma)} + \left({\frac{\rho}{\rho_\parallel}}\right)^2 }
\end{equation}

where, $L_{max}$ is the maximum length scale of the turbulence, c is the speed of light and the quantities $N_\parallel$ , $\gamma$ and $\rho_\parallel$ are constants specific to different kinds of turbulence. We have used fixed $L_{max} = 10^6$ km \cite{subramanian2009}. The quantity $\rho$ is expressed in terms of rigidity $R_g$ as,
\begin{equation}
\rho = \frac{R_L}{L_{max}} = \frac{R_g}{B_0 L_{max}}
\end{equation}
$\rho$ indicates how tightly cosmic rays are bounded by CME/magnetic cloud or shock.  
Here, for the CME-only model, $B_0$ refers to the magnetic field bounding the CME, and for the shock-only model, it refers to the enhanced magnetic field at the shock, and $R_L$ is the Larmor radius. 

The  $\sigma^2$ represents turbulence level which is given as,
\begin{equation}
\sigma^2 = \frac{\langle B_r ^2\rangle}{B_0 ^2}
\end{equation}

where $B_r$ is the turbulent magnetic field. 
The cross-field diffusion coefficient ($D_\perp$) is related with parallel diffusion coefficient $(D_\parallel)$ as

\begin{equation}
\frac{D_\perp}{D_\parallel} = \begin{cases} N_\perp (\sigma ^2)^{a_\perp}, & {\rho \leq 0.2 }  \\ N_\perp (\sigma ^2)^{a_\perp} (\frac{\rho}{0.2})^{-2}, & {\rho > 0.2 } \end{cases}
\end{equation}

The quantities $N_\perp$ and $a_\perp$ are constants specific to different kinds of turbulent spectra.   
For Kolmogorov turbulence $N_\parallel = 1.7$ , $\gamma = 5/3$, $\rho_\parallel = 0.20$, $N_\perp   = 0.025$ and $a_\perp = 1.36$.



\subsection{CME-only Model}
CME is considered to be a flux rope structure, propagating and expanding outwards from the Sun due to which shock is generated ahead of it. (see Figure 1 of \cite{Babu_et_al_2013} for a cartoon). The sheath region between flux rope and the shock is turbulent which is primary contributor in modulation of cosmic rays. We have used CME-only model elaborately discussed by \cite{subramanian2009} and here we briefly highlight the basic features.The magnitude of FD is a difference between the cosmic ray proton density inside and outside the CME/magnetic cloud that intercepts the Earth. The proton flux F ($cm^{-2} sec^{-1}$)  diffusing into the CME/magnetic cloud at a given time depends on the $D_\perp$ and the density gradient $\frac{\partial N_a}{\partial r}$, which is written as,

\begin{equation}
 F = D_\perp  \frac{\partial N_a}{\partial r}
\end{equation}

where, $N_a$ is the ambient density of high energy protons. It has been assumed that there are no galactic cosmic rays inside the CME when it ejects from the Sun. The ambient cosmic rays diffuse into the CME/magnetic cloud through out its propagation and expansion. Therefore, the total number of galactic cosmic rays diffused into the CME after time T can be calculated as,

\begin{equation}
 U_i = \int_0^T A(t) F(t) dt = \int_0^T D_\perp A(t) \frac{\partial N_a}{\partial r} dt
\end{equation}

The integration extends from the time, the CME is first observed near the Sun (t = 0) till the time (t = T ) when it reached at the Earth. CME/magnetic cloud is generally assumed as an expanding cylindrical flux rope whose length and cross sectional area increases as it propagates outward. A(t) is the cross-sectional area of the expanding CME/magnetic cloud which can be express as 

\begin{equation}
 A(t) = 2 \pi R(t) L(t)
\end{equation}

where, R(t) is the cross sectional radius and L(t) is the length of the cylindrical flux rope of CME/magnetic cloud at time t. Length L(t) of cylindrical flux rope  as can be approximated as,

\begin{equation}
 L(t) = \pi H(t) = \pi ( V t + R_0)
\end{equation}

here, H(t) is the height  of the cylindrical flux rope above the solar limb, V is the radial speed of of CME and typically 0.88 times the expansion speed ($V_{exp}$) of halo CMEs. Note that $R_0$ is the initial value of the observed height H of the flux rope above the solar limb. Generally, this value is considered to be of the order of few solar radii. The ambient density gradient $\frac{\partial N_a}{\partial r}$ can be approximated as,

\begin{equation}
 \frac{\partial N_a}{\partial r} \approx \frac{N_a}{R(t)}
\end{equation}

Therefore, by using all above equations we can estimate number of diffused cosmic rays inside the flux rope as,

\begin{equation}
 U_i = 4 \pi^2 N_a \int_0^T D_{\perp} H(t) dt
\end{equation}

The cosmic ray density inside the flux rope when it reaches near the Earth would be,

\begin{equation}
 N_i = \frac{U_i}{\pi R(T)^2 L(T)}
\end{equation}

where, R(T) and L(T) indicate the cross sectional radius and length of the cylindrical flux rope of CME/magnetic cloud near the Earth at time $`T'$. Therefore, this can be related to the the magnitude $M_{CME}$ of the Forbush decrease as,

\begin{equation}
 M_{CME} = \frac{N_a -N_i}{N_a} = \frac{\Delta N}{N_a} = 1 - \frac{4 \pi \int_0^T D_{\perp} H(t) dt}{R(T)^2 L(T)}
\end{equation}

It is important to note that, \cite{subramanian2009} have used quantity $ \alpha$ which actually denotes the fraction of the total decrease that can be attributed to CME/magnetic cloud. We have not considered this quantity in above expression of FD magnitude since, we have separately estimated the FD magnitude contribution of shock and CME.

\begin{figure}[tbp]
\begin{center}
\includegraphics[width = 10 cm]{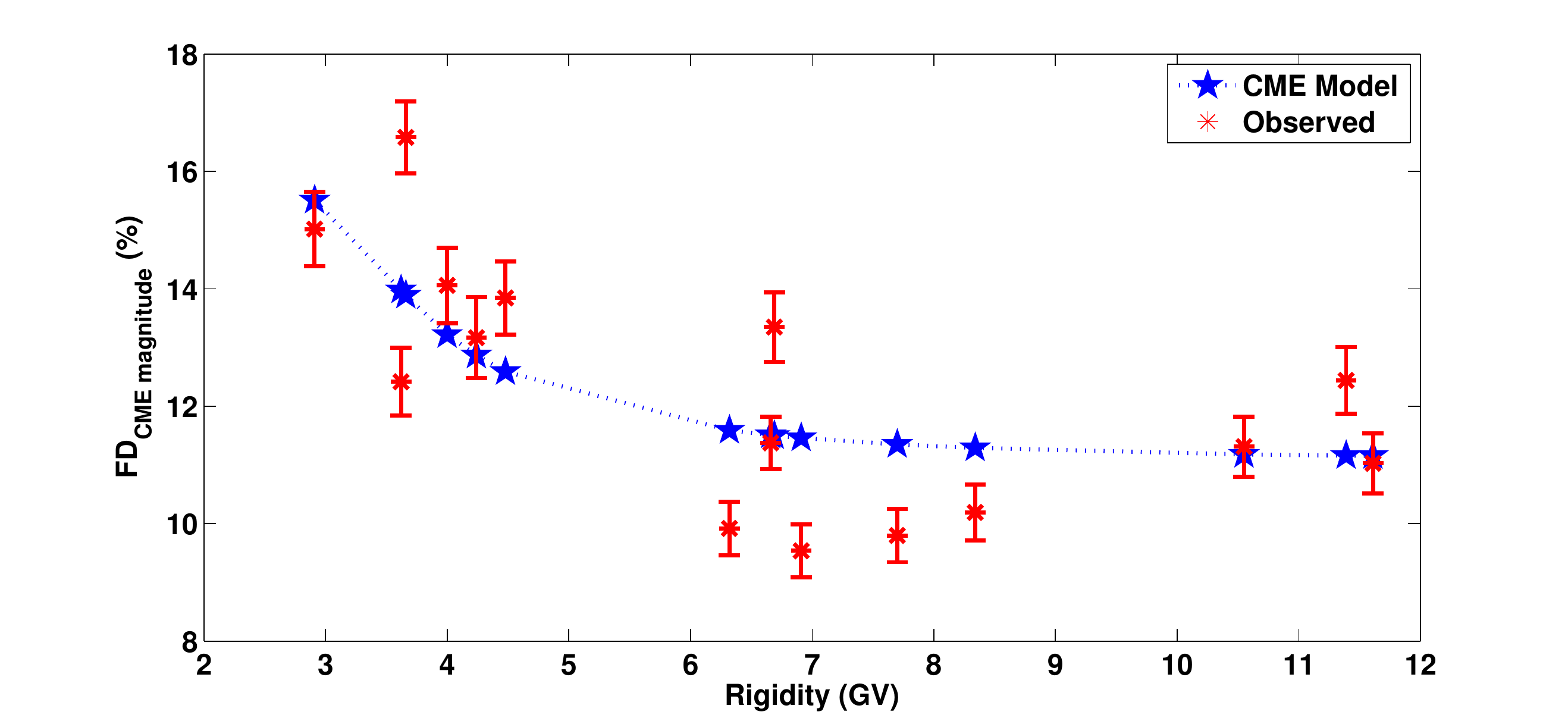}
\caption{Rigidity dependence of FD}
\label{fig:cme-model}
\end{center}
\end{figure}

To reproduce the observed FD amplitude corresponding to each given rigidity we have used this CME-only model. Since, event had occurred in 1978, no space based corona-graph observation were available. Eventhough, no interplanetary measurements were continuously available. This put constraint on us to use parameters from earlier reports on this event. We have used $ B_0 = 15 nT$  at $1 AU$ which is adopted from Geranios at el (1983) \cite{Geranios1983}. Total propagation time duration of CME from the Sun to the Earth was estimated $\sim$ 45.26 hours based on reported solar flare occurrence time and SSC. We have estimated radius of flux rope of CME/magnetic cloud from observed FD profile. As noted earlier, total time duration of the symmetric profile of FD is the transit time of CME. Therefore, the approximate radius of the flux rope is product of main phase duration and solar wind speed which turns out to be $1.18 * 10^{10} m$. The Model requires length scale of turbulence which we have adopted as $10^9 m$ \cite{subramanian2009}. Generally, the height of the flux rope just above the solar limb is few solar radii, therefore we have taken as $3.5 * 10^9 $ m ($\sim$ 5 solar radii). However, note that model is not very sensitive for this parameter. We have considered range of CME radial velocity as $1500 - 2500 km/s$ and for both limiting values, we have fitted CME-only model by minimizing $\chi^2$ with $\sigma^2$ as free parameter. The used and estimated parameters for CME-only model are listed in Table \ref{tab:CME_details}. The estimated (blue) and observed main phase FD amplitude (red) based on CME only model for different rigidities is shown in Figure \ref{fig:cme-model}.

\begin{table}[H]
\caption{Summary of various  used and observationally derived parameters using CME-only model for the FD} 

\begin{tabular}{l l l l l}
\\[0.5ex]
\hline
\hline

Parameters	&Value	&Remark\\[0.5ex]
\hline
\hline
$T_{CME}$	&$45.2666 * 3600$ sec	&The Sun-Earth travel time in hours for the CME\\
$B_{0}$	&$15 * 10^{-9}$ T	&Magnetic field inside the magnetic cloud \\
C	&$3 * 10^8$ m/s	&Speed of light\\
$L_{max}$	&$10^9$ m	&maximum length scale of turbulence \\

$R(T)$	&$1.18 * 10^{10}$ m	&Cross sectional radius of flux rope of CME at the Earth\\
$L(T)$	&$ \pi * 1.5 * 10^{11}$ m	&Length of flux rope of CME at the Earth\\
$R_0$	&$35 * 10^8$ m	&Initial value of the observed height of the CME above the solar limb\\
$V$	&$1500 - 2500$ km/s	&Radial speed of the CME\\
$\sigma_{CME}^2$	&$169 - 42.25 \%$ 	& Turbulence level for CME-only model\\
$\chi_{shock}^2$	&$6.35 - 6.36 $ 	& minimum $\chi^2$ for CME-only model\\

\hline 
\end{tabular}
\label{tab:CME_details}
\end{table}

\subsection{Shock-only Model}

\begin{figure}[tbp]
\begin{center}
\includegraphics[width = 10 cm]{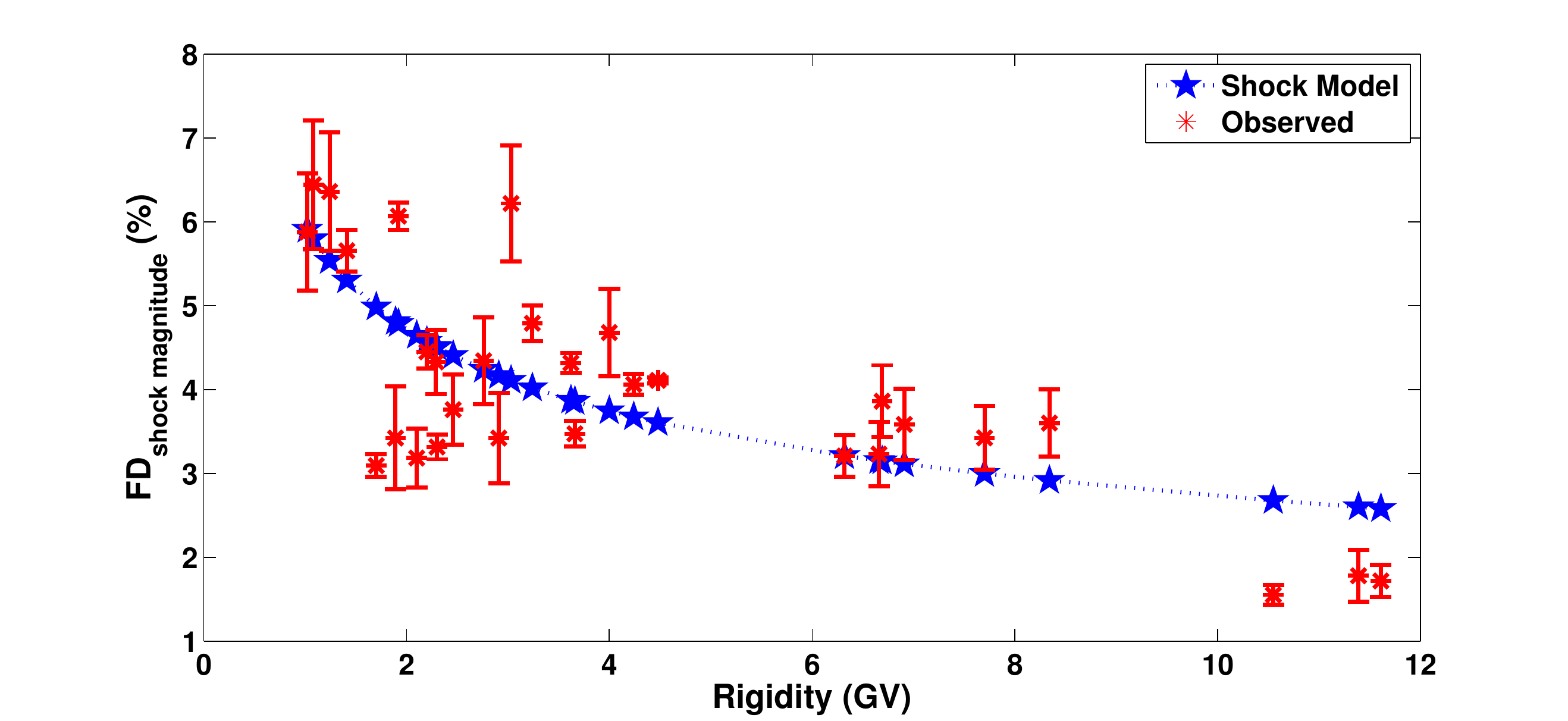}
\caption{Rigidity dependence of FD}
\label{fig:shock-model}
\end{center}
\end{figure}

Shock-only model is based on the diffusion of cosmic rays across the propagating diffusive barrier. Right behind the shock-front solar wind velocity and magnetic field enhances which can reduce the drift of cosmic rays. The changes in the magnetic topology behind the shock can give rise to reduce radial diffusion of cosmic rays. Along with this it is believed that region behind the shock front namely shock-sheath is turbulent in nature. Wibberenz 1998, had provided formalism of shock-only model by taking into account above mention physical scenario. The diffusion coefficient is suppressed inside the shock-sheath region and gives rise to radial variation of cosmic ray density. When the Earth passes through the shock region depression of cosmic rays flux is observed. The magnitude and duration of this depression depends on the properties of shock-sheath and ambient medium. Note that, shock only model needs diffusion coefficient of disturbed and undisturbed medium unlike CME-only model. The expression for the magnitude of the Forbush decrease according to this model is \cite{wibberenz1997, wibberenz1998, Babu_et_al_2013}
\begin{equation}
 M_{shock} = \frac{N_a - N_{shock}}{N_a} = \frac{\Delta N}{N_a} = \frac{V_{sw} L_{shock}}{D_\perp^a} (\frac{D_\perp^a}{D_\perp^{shock}} - 1)
\end{equation}

where $N_a$ and $N_{shock}$ are cosmic ray densities inside the ambient and shock-sheath medium respectively. $D_\perp^a$ and $D_\perp^{shock}$ are the perpendicular diffusion coefficients inside the ambient and shock-sheath medium respectively. $V_{sw}$ is the solar wind velocity and $L_{shock}$ is the shock sheath thickness. In computing $D_\perp^a$ and $D_\perp^{shock}$, we need to use different values for the proton rigidity $\rho$ for the ambient medium and in the shock sheath; they are related to the proton rigidity $R_g$ by
\begin{equation}
\rho^a = \frac{R_g}{B_0^a L_{shock}} 
\end{equation}

\begin{equation}
\rho^{shock} = \frac{R_g}{B_0^{shock} L_{shock}}
\end{equation}

where $B_0^a$ is the ambient magnetic field, $B_0^{shock}$ is the magnetic field inside the shock sheath.
We have estimated shock phase amplitude of FD by using shock only model for each given rigidity. The shock thickness is determine by using $D_{st}$ index and neutron monitor data. The main phase of geomagnetic storm started after 10 hours  of SSC onset. Also, shock phase duration of the FD last for $\sim$ 10 hours. This duration roughly represents shock-sheath transit time at the Earth. The shock-front speed adopted as $900 km/sec$ \cite{wada1979}. We have estimated sheath thickness which is product of shock-front speed and shock phase duration.    The ambient interplanetary magnetic field and enhanced magnetic field inside the shock are adopted from Geranios (1983).  The shock only model is fitted by reducing $\chi^2$ with $\sigma^2$ as free parameter. The various parameters used and derived using shock-only model are listed in Table \ref{tab:shock_details}. The estimated (blue) and observed shock phase FD amplitude (red) based on shock-only model for different rigidities is shown in Figure \ref{fig:shock-model}.


 \begin{table}[H]
\caption{Summary of various  used and observationally derived parameters using shock-only model for the FD} 

\begin{tabular}{l l l l l}
\\[0.5ex]
\hline
\hline
Parameters	&Value	&Remark\\[0.5ex]
\hline
\hline

$V_{sw}$	&$900$ km/sec	&Shock front speed near Earth \\
$T_{shock}$	&$10 * 3600$ sec	& Duration of shock \\
$B_a$	&$5 * 10^{-9}$ T	&magnetic field in the ambient solar wind\\
$B_{shock}$	&$15 * 10^{-9}$ T	&magnetic field in the shock \\
C	&$3 * 10^8$ m/s	&Speed of light\\
$L_{max}$	&$1.5 * 10^{11}$ m	&maximum length scale of shock \\
$\sigma_{shock}^2$	&$12.25 \%$ 	& Turbulance level for shock-only model\\
$\chi_{shock}^2$	&$13.85 $ 	& minimum $\chi^2$ for shock-only model\\
\hline 
\end{tabular}
\label{tab:shock_details}
\end{table}

\section{Discussion and conclusions}

This is the unique event which exhibits explicitly shock phase, main phase, fast recovery phase and slow recovery phase. As noted earlier in Figure \ref{fig:average}, SSC and main phase of geomagnetic storm coincides with onset of shock phase and main phase of the FD respectively. This clearly demonstrates gradual decrease corresponds to the shock phase and slow component of recovery phase (assume CME part of FD is absent) indicates shock effect. However, the symmetric part of the FD (main phase and rapid recovery phase) are associated with passage of the Earth through CME. These clearly indicate that profile of FD give indirect information of the internal structure of CME and shock.

The rigidity response of FD shows temporal variation of the FD in different rigidity bands (see Figure \ref{fig:stack_rigi}). The main/shock phase amplitude of the FD decreases with increase in rigidity. After the maximum FD depression, normalized neutron flux remains steady for few hours in 2.2 to 8.5 GV cutoff rigidity band. This can be explained by the following physical scenario. 
The total diffusion coefficient for high energy cosmic rays is higher as compared to that of lower energy cosmic rays. The high diffusion coefficient allows high energy cosmic rays to penetrate deep inside the flux rope as compare to the low energy cosmic rays. 
This give rise to high density of cosmic rays of high energy inside the flux rope as compare to that of low energy cosmic rays. 
This implies higher FD amplitude for low energy cosmic rays as compare to that of high energy. This also indicates low radial gradient of cosmic ray density in CME for high energy cosmic rays. This reflects in slope of main phase and fast recovery phase for different rigidities. As CME propagates and evolve in interplanetary space cosmic ray density inside the CME kepdf increasing till it reaches ambient medium cosmic ray density for particular energy. Note that, high energy cosmic ray density inside the CME approaches ambient density earlier as compare to low energy cosmic rays. We are speculating that the efficient penetration of high energy particles results in development of finite region inside the CME having almost negligible gradient of high energy cosmic ray density. Therefore, there is a region in CME near the center of the flux rope having nearly constant flux of high energy diffused particles. Therefore, when the Earth passes through this extended minimum is observed by high cut-off rigidity neutron monitors. However, we have observed deviation in FD profile for 10 - 12 GV which need to be investigated in detail.

The power law behavior has been observed for both shock phase and total FD magnitude showing almost same power law index. 
It is interesting to see almost same value of power index for both the amplitudes. It might indicate common physical mechanism characterized by the power law index.  The almost same power index might be related to the turbulence in magnetic field associated with shock and ejecta.

The average ratio of shock to main phase amplitude in different rigidities is $\sim$5, which indicates the decrease associated with the shock has about $20 \%$ contribution in the total amplitude of the FD. The shock phase amplitude shows local time variation for rigidities $< 3.1$ GV. However, no clear local time variation is observed for the main phase amplitude of FD. The shock phase amplitude is lowest in the evening-night sector and maximum in the dawn sector. Also, it appears that amplitude gets saturated in the dusk-night sector. The local time variation of maximum depression time (in UT) of main phase revealed interesting observations. The maximum depression time of the main phase in evening-night sector is simultaneous. The main phase reached to its maximum depression earliest at dawn and get delayed towards afternoon local time sector. The afternoon local time sector showed delay of approximately 6 hours as compared to the dawn sector. These observations are analogues to the local time variation of shock phase amplitude (see Figure \ref{fig:LT_shock}). This local time dependence of shock phase amplitude and maximum depression time might indicate orientation effect of the interplanetary shock and CME. It has been reported by Garnios et. al. (1983) that the shock associated with this event incident at dawn sector i.e perpendicular incidence at the dawn \cite{Geranios1983}. However, to understand shock/CME orientation effect needs more detail investigation. 

FD rapid recovery followed the main phase of FD and then slow component of recovery continued for few days. There is no unambiguous local time dependence observed for estimated recovery time constants for fast recovery phase ($\tau_{1}$). Note that, slow component has recovery time constant ($\tau_{2}$) of the order of the day as compare to ($\tau_{1}$) which is approximately few hours, so its local time variation is not relevant to study. The rigidity dependence of $\tau_{1}$ does not show any clear picture. This might be due to  it's association with the ejecta/flux rope. However $\tau_{2}$ shows a clear relationship with the rigidity. It remains constant for $0.5-4$ GV and decreases for $4-12$ GV.  This indicates the slow component recovers faster in higher rigidity as compared to that of lower rigidity. The recovery time depends on the diffusion coefficient and refilling of the interplanetary space swept by the CME/shock. The short recovery time constant of slow recovery component for higher rigidity indicates high rigidity cosmic rays easily diffuse into the interplanetary medium as compared to the low rigidity cosmic rays. Also note that slow recovery correlates well with corresponding solar wind velocity decrease measured by Helios 1 and 2. The FD slow recovery phase better correlates with Helios 1 as compared to the Helios 2. This might be due to the difference in position of both the spacecrafts, Helios 1 was closer to the ecliptic as compared to the Helios 2. The detail study of solar wind correlation with FD recovery will be reported elsewhere.

As described earlier, in present study we have used two basic models which are mainly based on diffusion of cosmic rays through the ordered and compressed large scale magnetic field of flux rope/shock. However, turbulence corresponding to flux rope/shock affects the diffusion of cosmic rays. The modulation of cross field and parallel diffusion of cosmic rays due to turbulence is the basis of these models. In earlier work of Subramanian et al. 2009 and Arunbabu at al. 2013, it has been demonstrated that FD amplitude is well estimated using CME only model. However, the FD main phase is mainly dominated by ejecta, so shock-only model can not properly estimate FD amplitude by using realistic turbulence values. Therefore, according to our knowledge, first time we have applied shock-only and CME-only model separately to estimate turbulence level required for observed shock phase amplitude and main (CME) phase amplitude respectively. Shock only model estimated observed shock phase amplitude for turbulence level $\sigma^2 = 12.25 \%$ with minimum $\chi^2  = 13.85$. Similarly, CME only model estimated observed main phase amplitude for turbulence level  $\sigma^2 = 42.25 \%$ with minimum $\chi^2  = 6.36$ considering radial speed of CME $V = 2500 km/sec$. Turbulence level is free parameter in model fitting by which we have estimated the required magnetic energy in the turbulence to explain the observed FD amplitude for both shock and CME. The estimated turbulence level appears to be realistic for the studied event.

This study demonstrate and emphasize that the CME-only and shock-only model must be applied to main phase and shock phase FD amplitude separately. This confirms present accepted physical scenario that the first step of FD is due to propagating shock barrier and second step is due to flux rope of CME/magnetic cloud.


\acknowledgments

We are thankful to World Data Center in Russia and Ukraine for making neutron monitor observatories data available. We are also thankful to all neutron monitor observatories listed in Table \ref{tab:lab_details}. We are thankful to WDC and CDAWeb for making geomagnetic data and solar wind data available. We are thankful to Department of Physics, University of Mumbai, Mumbai and Indian Institute of Geomagnetism, Panvel, Navi-Mumbai for providing resources and facilities during data analysis. We extend our heartfelt thanks to Dr. M. R. Press, N. Bijewar of Department of Physics for important discussion and encouragement. We would like to express our gratitude to Deva Nandan and Sreeba of Indian Institute of Geomagnetism, Navi Mumbai. 









\end{document}